# Interstellar Conditions Deduced from Interstellar Neutral Helium Observed by IBEX and Global Heliosphere Modeling

P. Swaczyna[1,2,*], M. Bzowski[2], J. Heerikhuisen[3], M. A. Kubiak[2], F. Rahmanifard[4], E. J. Zirnstein[1], S. A. Fuselier[5,6], A. Galli[7], D. J. McComas[1], E. Möbius[4], N. A. Schwadron[1,4]

[1]Department of Astrophysical Sciences, Princeton University, Princeton, NJ 08544, USA
[2]Space Research Centre PAS (CBK PAN), Bartycka 18a, 00-716 Warsaw, Poland
[3]Department of Mathematics and Statistics, University of Waikato, Hamilton, New Zealand
[4]Physics Department, Space Science Center, University of New Hampshire, Durham, NH 03824, USA
[5]Southwest Research Institute, San Antonio, TX 78228, USA
[6]University of Texas at San Antonio, San Antonio, TX 78249, USA
[7]Physics Institute, University of Bern, Bern, 3012, Switzerland

## Abstract

In situ observations of interstellar neutral (ISN) helium atoms by the IBEX-Lo instrument onboard the Interstellar Boundary Explorer (IBEX) mission are used to determine the velocity and temperature of the pristine very local interstellar medium (VLISM). Most ISN helium atoms penetrating the heliosphere, known as the primary population, originate in the pristine VLISM. As the primary atoms travel through the outer heliosheath, they charge exchange with He$^+$ ions in slowed and compressed plasma creating the secondary population. With more than 2.4 million ISN helium atoms sampled by IBEX during ISN seasons 2009-2020, we compare the observations with predictions of a parametrized model of ISN helium transport in the heliosphere. We account for the filtration of ISN helium atoms at the heliospheric boundaries by charge exchange and elastic collisions. We examine the sensitivity of the ISN helium fluxes to the interstellar conditions described by the pristine VLISM velocity, temperature, magnetic field, and composition. We show that comprehensive modeling of the filtration processes is critical for interpreting ISN helium observations, as the change in the derived VLISM conditions exceeds the statistical uncertainties when accounting for these effects. The pristine VLISM parameters found by this analysis are the flow speed (26.6 km s$^{-1}$), inflow direction in ecliptic coordinates (255.7°, 5.04°), temperature (7350 K), and B-V plane inclination to the ecliptic plane (53.7°). The derived pristine VLISM He$^+$ density is 9.7×10$^{-3}$ cm$^{-3}$. Additionally, we show a strong correlation between the interstellar plasma density and magnetic field strength deduced from these observations.

## 1. Introduction

The motion of the Sun through the very local interstellar medium (VLISM) results in an inflow of interstellar neutral (ISN) atoms, also known as the interstellar wind, to the heliosphere (Patterson et al. 1963; Fahr 1968). ISN atoms are ionized inside the heliosphere by charge exchange with solar wind ions, photoionization, and electron impact ionization (Bzowski et al. 2013). The combined ionization rates for hydrogen are higher than those for helium atoms, and therefore the ISN hydrogen density at 1 au is at most ~5% of their density at the termination shock (Ruciński & Bzowski 1995). Consequently, while more than 10 times less abundant than ISN hydrogen atoms in the VLISM, ISN helium atoms, which have survival probabilities of ~65% during similar solar cycle phases (Swaczyna et al. 2022a), become the main ISN population at 1 au. Observations of ISN helium inside the heliosphere allow for finding the interstellar conditions in the VLISM near the Sun (Möbius et al. 2004).

The IBEX-Lo instrument (Fuselier et al. 2009) on the Interstellar Boundary Explorer (IBEX, McComas et al. 2009) provides high-statistics observations of ISN atoms (Möbius et al. 2009a, 2009b). Despite the narrow energy distribution of ISN helium atoms in the solar frame (Sokół et al. 2015), their observations span over the four lowest electrostatic analyzer (ESA) steps (Swaczyna et al. 2018; Galli et al. 2022). This broad energy distribution occurs because helium atoms sputter a broad energy distribution of negative ions (mostly H⁻) off of the instrument's conversion surface. Comparison of the observed count rates in different energy steps allows for the determination of the relative response of the instrument (Schwadron et al. 2022; Swaczyna et al. 2023a). The IBEX-Lo observations of ISN helium atoms have been analyzed in a series of papers to find the flow vector and temperature of the VLISM (Bzowski et al. 2012, 2015; Möbius et al. 2012, 2015; Leonard et al. 2015; Schwadron et al. 2015; Swaczyna et al. 2018, 2022a).

---

[*] Corresponding author (pswaczyna@cbk.waw.pl)

The studies aiming to find the VLISM flow vector and temperature from IBEX observations need to account for various effects that modify the ISN helium population near the heliospheric boundaries. They must account for helium atom ionization inside the heliosphere and solar gravity that deflects their trajectories (Lee et al. 2012, 2015; Sokół et al. 2015). Still, early studies of IBEX data did not account for any filtration processes operating beyond the heliopause or other ISN helium populations and thus assumed a single Maxwell distribution outside the heliopause (Bzowski et al. 2012; Möbius et al. 2012). These analyses concluded that the speed and flow direction significantly differ from those estimated from the Ulysses mission (Witte et al. 2004). This discrepancy appeared statistically significant and suggested that the flow velocity changes over time (Frisch et al. 2013, 2015), but this conclusion was controversial (Lallement & Bertaux 2014). Eventually, an additional population found in the IBEX data in follow-up studies, dubbed the Warm Breeze, was critical to explain this discrepancy (Kubiak et al. 2016, 2014). Later IBEX data analyses, which accounted for the Warm Breeze (Bzowski et al. 2015; McComas et al. 2015a, 2015b), showed that the derived VLISM parameters agree with those obtained from reanalyses of Ulysses observations (Bzowski et al. 2014; Wood et al. 2015).

The Warm Breeze flow velocity is from a direction deflected away from the inflow direction of the pristine VLISM flow along the so-called B-V plane defined by the interstellar flow and magnetic field directions (Kubiak et al. 2016). This coplanarity indicates that the Warm Breeze is the secondary population of ISN helium atoms created by neutralized He⁺ ions in the outer heliosheath, where the interstellar plasma slows down in front of the heliopause. Indeed, further simulations of the secondary ISN helium production confirmed that the Warm Breeze is produced in this process (Bzowski et al. 2017). Subsequently, Bzowski et al. (2019) used the relative abundance of the Warm Breeze to estimate the He⁺ ion density in the VLISM.

In addition to charge exchange collisions removing the primary population and producing the secondary population, elastic collisions may lead to angular scattering of both populations and, hence, to the momentum transfer between charged and neutral populations in the VLISM. Swaczyna et al. (2021) showed that while only a small fraction of primary ISN helium atoms charge exchange and produce a secondary atom, these atoms elastically collide with protons, He⁺ ions, and hydrogen atoms typically ~4 times while traveling through the outer heliosheath. These collisions slow down and heat the primary population by ~0.5 km s$^{-1}$ and ~800 K, respectively (Swaczyna et al. 2023b).

The filtration of the ISN helium populations manifested in the amount of deflection of the Warm Breeze from the pristine VLISM flow vector reflects the global structure of the heliosphere. This structure depends on interstellar conditions, including the flow vector, temperature, composition, and magnetic field (e.g., Pogorelov et al. 2017). Therefore, the filtered ISN helium atoms observed by IBEX are sensitive to these parameters. This paper presents the first analysis aiming to use the observations of filtered ISN helium atoms from IBEX to constrain simultaneously the set of interstellar properties such as the flow velocity, temperature, composition, and magnetic field vector.

## 2. IBEX-Lo Data

We use IBEX-Lo ISN helium observations in energy steps 1–3, covering observations from ISN seasons 2009-2020 as analyzed by Swaczyna et al. (2022a)[†]. The instrument boresight follows a great circle perpendicular to the spacecraft's spin axis. The onboard computer accumulates the IBEX-Lo observations into 6° histogram bins in the spacecraft spin angle. We use the bins for the spin angle centers from 216° to 318°, i.e., around the ecliptic plane located at an angle of 270°. The observations are accumulated over time when the spacecraft spin axis is kept constant. The spacecraft spin axis is adjusted every few days and approximately follows the Sun. We include data collected when the ecliptic longitude of the spin axis pointing is between 235° and 335°. These ranges are broader than those used in the primary ISN analysis focusing only on the primary population peak (Bzowski et al. 2015; Swaczyna et al. 2018, 2022a). We calculate the count rate for each data point, apply throughput correction, and subtract the ubiquitous background rate (Galli et al. 2015, 2017). The throughput correction is needed to compensate for losses of observed events due to processing a high volume of electron background events transmitted to the onboard computer in seasons 2009-2012 (Swaczyna et al. 2015). The change to the time-of-flight section logic starting with ISN seasons 2013 eliminated these losses.

Due to the high number of observed counts during the ISN seasons, the uncertainty analysis for ISN observations must also include a few systematic uncertainties. Swaczyna et al. (2015, 2018, 2022a) performed a detailed uncertainty analysis, including contributions from throughput correction, background, spin axis pointing, spin pulse accuracy, and spin angle offset. In the current study, we follow their analysis, except we do not have uncertainties related to the Warm Breeze parameters because in contrast to the previous work the secondary population is now part of our modeling. The uncertainty matrix accounts for correlations introduced by the systematic source of uncertainty.

Figure 1 presents the ISN helium observations in ESA 1–3 used in this analysis compared with the modeled count rate (see Section 3). The count rates are accumulated into 6°×6° pixels based on the mean instrument boresight pointing for each data point. The figure presents the accumulated data spanning ISN seasons from 2009-2020. The count rates corresponding to data points obtained in the ISN season 2013 and later are multiplied by a factor of 1.96, which compensates for the reduction in the instrument efficiency after the post-acceleration voltage was lowered from the nominal 16 kV to 7 kV. We find this factor from the normalization norms discussed in Section 3. The maps show that the ISN signal near the peak increases from ESA step 1 to 3. However, the rates are reduced in ESA 3 at larger ecliptic latitudes because the atoms' speeds in the spacecraft frame are too low to be registered in this energy step. Detailed discussion of the IBEX-Lo relative response for ISN helium can be found in Swaczyna et al. (2023a). The peak of the ISN helium signal is shifted away from the flow velocity (marked as V in the figure) because solar gravity deflects ISN helium atoms' trajectories. We exclude the data within ~10° of the pristine flow direction because they include a significant contribution from ISN hydrogen (Saul et al. 2012; Galli et al. 2019; Rahmanifard et al. 2019). Hydrogen atoms converted to $H^-$ ions at the conversion surface are observed in ESA steps 1 and 2, and overlap with the ISN helium signal.

---

[†] The derived data products are available on Zenodo: https://doi.org/10.5281/zenodo.5842421; the raw products are available on the IBEX website: https://ibex.princeton.edu/RawDataReleases.

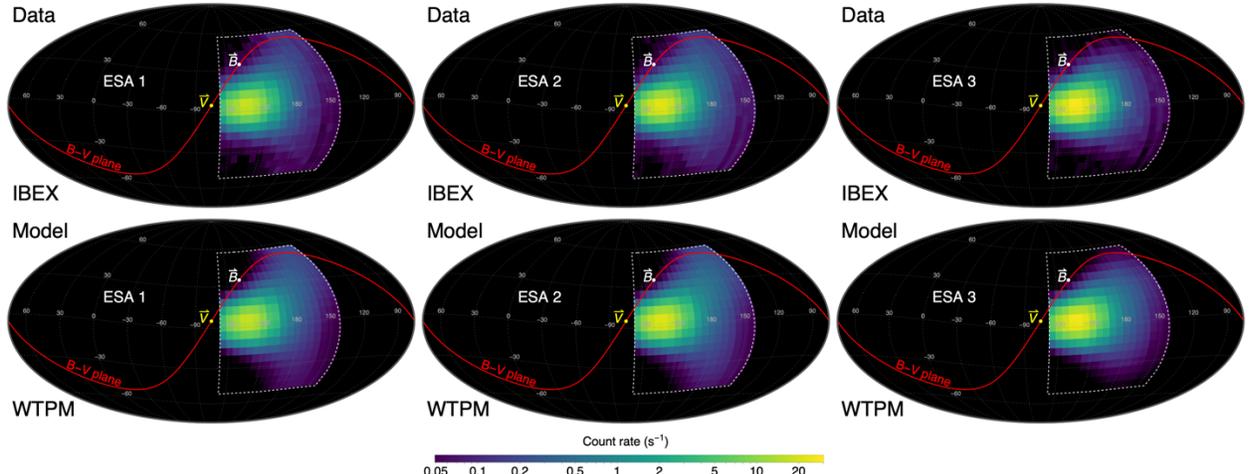

**Figure 1.** IBEX-Lo observations (top row) of the ISN helium in ESA steps 1–3 (left to right) compared with the modeled count rates (bottom row) as projected in the sky based on the instrument boresight pointing. The pristine VLISM flow direction ($\vec{V}$, yellow dot) and magnetic field ($\vec{B}$, white dot) are shown together with the B-V plane (red line) defined by these two directions. All maps use the same color scale. The count rates near the peak slightly increase from ESA 1 to 3, but the rates are significantly reduced at higher latitudes in ESA 3.

## 3. Modeling the ISN Helium Flux

To find the pristine VLISM conditions, we quantitatively compare the observed count rates with the model ISN helium flux integrated over the instrument response and the observation time for each data point defined by the spin axis pointing, spin angle bin, and ESA step. Note that we do not use the accumulated observations as shown in Figure 1 but the corrected count rates in each data point separately. We employ two integration schemes for this modeling: the Warsaw test particle model (WTPM, Sokół et al. 2015) and the analytic full integration model (aFINM, Schwadron et al. 2015). Two closely collaborating teams independently developed these models, and the results were compared by Möbius et al. (2015). The main difference between these approaches is that the WTPM solves the Kepler trajectories for all intermediate points between the VLISM and the detector, integrating the ionization losses using a time-dependent ionization model. By contrast, the aFINM uses a simplified approach in which the ionization rate is assumed to be constant in time along the trajectory.

Both models were originally developed to track atoms beyond the heliopause, where the phase space density was obtained from an assumed Maxwell distribution. However, Bzowski et al. (2017, 2019) expanded the WTPM to calculate changes to the statistical weights due to charge exchange collisions occurring in the outer heliosheath. The methodology requires flows and temperature of the plasma component in the outer heliosheath. The charge exchange losses and gains to the ISN helium population are calculated along each Keplerian trajectory. Similarly, the aFINM was generalized to calculate the ISN helium fluxes at IBEX from numerical distribution functions defined at 100 au from the Sun (Swaczyna et al. 2023b). The distribution at 100 au is calculated by solving the transport equation with the loss and gain terms using Monte Carlo integration. This methodology includes the angular scattering of ISN helium velocities in elastic and charge exchange collisions leading to momentum transfer and requires an external model to provide flow and temperature distributions in the outer heliosheath.

The flows and temperatures in the outer heliosheath for this study are taken from a global hybrid heliosphere simulation employing magnetohydrodynamic description of plasma and kinetic description of neutrals (Zirnstein et al. 2016; Heerikhuisen et al. 2019). The inner boundary conditions of the model at 1 au are selected based on the mean solar wind speed and density recorded in the OMNI database during four solar cycles before the maximum of solar cycle 24: plasma density including 4% of α particles is 8.3 nuc cm$^{-3}$, bulk speed – 441.5 km s$^{-1}$, and temperature – 51,000 K. The radial component of the magnetic field is 37.5 µG. We use four solar cycles to estimate these parameters because the transport of ISN helium atoms through the heliosphere takes several decades (Bzowski & Kubiak 2020). The global heliosphere model for our calculations should represent the outer heliosheath conditions at the times the atoms detected by IBEX over the last 12 years traversed the outer heliosheath. Importantly, the most recent solar cycles show significantly weaker solar wind (McComas et al. 2013). Moreover, because the model uses constant boundary conditions and does not account for the time evolution of the outer heliosheath, we average them over such a long period. Additionally, Fraternale et al. (2023) noted that the solar wind density at 1 au needs to be increased by 23% compared to the density derived from OMNI to match the density observed by SWAP on New Horizons (Elliott et al. 2019; McComas et al. 2021). However, the explanation for this discrepancy is beyond the scope of this paper.

The sensitivity study requires a pair of values for each of the parameters characterizing the outer boundary, i.e., a baseline and a modified value, as shown in Table 1. In our study, we compare the observed count rates with linear expansion of modeled count rates spanned by two models with the baseline and modified value in each considered parameter. The pristine VLISM velocity and temperature baseline values are taken from McComas et al. (2015b), which are widely used in other studies. Their variations (differences between the modified and baseline values) correspond to the uncertainties obtained by Bzowski et al. (2015), but the signs of the variations are selected to better match the results by Swaczyna et al. (2018). The pristine VLISM magnetic field vector is adopted based on the analysis of the IBEX ribbon position by Zirnstein et al. (2016). The magnetic field direction is defined by the inclination of the B-V plane to the ecliptic plane ($\gamma$) and the angle between the inflow vector and magnetic field (B-V angle, $\alpha$). Because we expect that the ISN helium does not constrain the magnetic field as strongly, we used variations equal to the uncertainty reported in their paper multiplied by a factor of 3. We chose the sign of this variation to represent a stronger magnetic field with a smaller B-V angle as obtained in some other studies, e.g., Izmodenov & Alexashov (2015).

The baseline plasma density in the pristine VLISM is adopted as 0.0856 cm$^{-3}$ to reproduce the Voyager 1 heliopause crossing under recent weaker solar wind conditions. The ISN hydrogen density is assumed to be 0.11 cm$^{-3}$. We chose this value to match the termination shock density of ISN hydrogen of 0.09 cm$^{-3}$ (Bzowski et al. 2009). We selected this value before the new estimate of this density from New Horizons was published (Swaczyna et al. 2020). The modified plasma and ISN hydrogen densities are from Bzowski & Heerikhuisen (2020). The model for these values gives the termination shock ISN hydrogen density closer to the New Horizons estimate. In this study, we vary the parameters around the baseline conditions to check the impact of the VLISM parameters on the IBEX ISN helium observations. Therefore, the selection of specific baseline values is not critical as the VLISM parameter may vary from the baseline values.

**Table 1. Initial Pristine VLISM Conditions for Modeling**

| Parameter | Symbol $p$ | Baseline value $p^0$ | Variation $\Delta p$ | Modified value $p^1$ |
|---|---|---|---|---|
| Speed | $v$ (km s$^{-1}$) | 25.4 | +0.4 | 25.8 |
| Inflow ecliptic longitude | $\lambda$ (°) | 255.7 | –0.5 | 255.2 |
| Inflow ecliptic latitude | $\beta$ (°) | 5.1 | –0.1 | 5.0 |
| Temperature | $T$ (K) | 7500 | +260 | 7760 |
| ISN hydrogen density | $n_{H^0}$ (cm$^{-3}$) | 0.11 | +0.044 | 0.154 |
| Plasma density | $n_{pl}$ (cm$^{-3}$) | 0.0856 | –0.0106 | 0.075 |
| Magnetic field strength | $B$ (µG) | 2.93 | +0.24 | 3.17 |
| B-V angle | $\alpha$ (°) | 39.5 | –1.8 | 37.7 |
| B-V plane inclination | $\gamma$ (°) | 52.2 | +3.6 | 55.8 |
| He$^+$ density | $n_{He^+}$ (cm$^{-3}$) | 0.00898 | +0.00036 | 0.00934 |

The global heliosphere model used in our study does not include He$^+$ ions as a separate fluid. Instead, we assume that the total plasma density is distributed between protons and He$^+$ ions in constant proportions throughout the outer heliosheath, assuming that $n_{\text{pl}} = n_{\text{p}} + 4n_{\text{He}^+}$, where we scale the He$^+$ ion contribution by their mass. The baseline pristine VLISM density of He$^+$ is assumed to be 8.98×10$^{-3}$ cm$^{-3}$, as found by Bzowski et al. (2019). For the modified values, we use the central value plus 3σ uncertainty. Furthermore, we assume that both populations are in equilibrium, i.e., that they have the same bulk velocity and temperature. This procedure is only aimed at estimating the He$^+$ ion properties. While Fraternale et al. (2021) found that the protons and He$^+$ have the same temperature due to Coulomb collisions, both the presence of He$^+$ ions and accounting for electrons as a separate fluid change the temperature and the heliosphere structure (Fraternale et al. 2023). Because the model we use does not self-consistently account for these effects, they may be a source of systematic uncertainties in our analysis.

We model the ISN helium fluxes with different interstellar parameters. We separately estimate the relative response function for the integrated flux at IBEX in each considered case (Swaczyna et al. 2023a). The obtained response functions are similar in each case, with the largest difference expected for the variation in speed. The obtained relative response is applied to each data point. Additionally, the simulated flux is also multiplied by a factor:

$$S_i(\boldsymbol{\pi}) = a_{\text{year}(i)}\big(1 + b(v_i(\boldsymbol{\pi}) - v_0)\big), \tag{1}$$

where $i$ enumerates the data points in the study, $v_i$ is the mean ISN atom speed at IBEX, $v_0 = 78$ km s$^{-1}$ is the reference speed, $\boldsymbol{\pi}$ represents the vector of considered interstellar parameters, $a_{\text{year}(i)}$ is a normalization factor in cm$^2$sr, separate for each observation season, and $b$ (in km$^{-1}$s) describes possible common response function. Swaczyna et al. (2022a) found that a separate normalization factors are needed to match the IBEX data either due to change in the instrument sensitivity over time or an incomplete ionization model. The parameters $a_{\text{year}}$ and $b$ are obtained from the minimization described in the next section. However, because the relative response between energy steps is already factored in using the relative response function as a function of energy (Swaczyna et al. 2023a), the parameters in Equation (1) are common for all considered ESA steps. The product of the factor given in Equation (1), the relative response function, and the simulated integrated flux is denoted as $g_i(\boldsymbol{\pi})$ and is called the model count rate.

We interpolate the integrated ISN helium flux in each IBEX data point from the calculations for various interstellar conditions using the following series expansion similar to Equation (6) in Swaczyna et al. (2022a):

$$g_i(\boldsymbol{\pi}) = g_i(\boldsymbol{\pi}^0) + \sum_p \frac{(p - p^0)}{\Delta p}\Big(g_i(\boldsymbol{\pi}_p^1) - g_i(\boldsymbol{\pi}^0)\Big) + \Big(g_i^{+\text{MT}}(\boldsymbol{\pi}^0) - g_i^{-\text{MT}}(\boldsymbol{\pi}^0)\Big). \tag{2}$$

The sum is calculated over the interstellar parameters $p$ listed in Table 1, $g_i(\boldsymbol{\pi}^0)$ is the model count rate for the baseline parameters, $g_i(\boldsymbol{\pi}_p^1)$ denotes the model count rate in which parameter $p$ is replaced with the modified value $p^1$ from Table 1. The integrated fluxes in this sum are calculated using the WTPM. The last bracket in Equation (2) represents the difference between the count rate modeled using the aFINM with the momentum transfer ($g_i^{+\text{MT}}(\boldsymbol{\pi}^0)$) and without the momentum transfer ($g_i^{-\text{MT}}(\boldsymbol{\pi}^0)$). We use two models because the WTPM can efficiently calculate several cases needed in the sum, while only the aFINM can currently account for the momentum transfer effects. Figure 2 shows these differences revealing that most changes in individual parameters result in distinctive patterns of the model count rate changes within the data range used here.

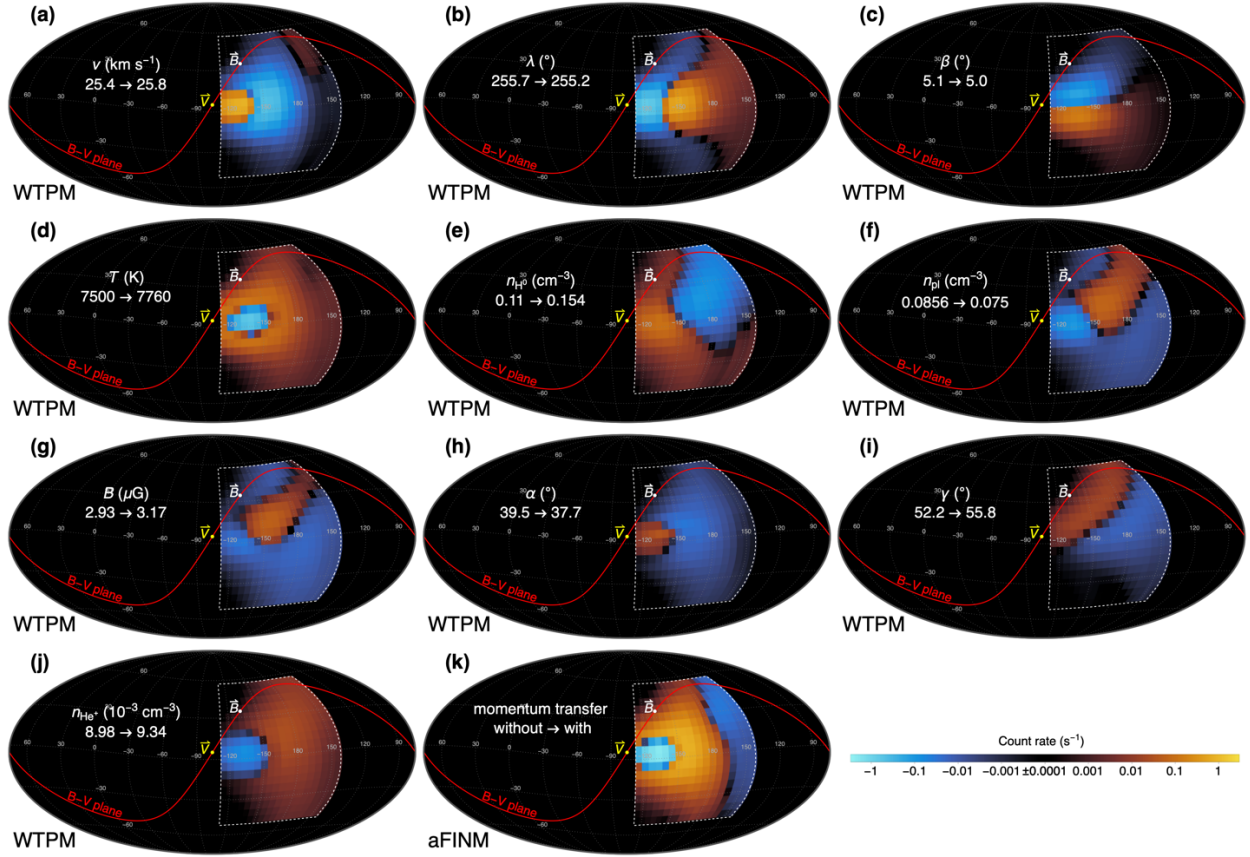

**Figure 2.** Changes to the model count rate as a result of modifications of interstellar parameters: (a) speed, (b) inflow longitude, (c) inflow latitude, (d) temperature, (e) hydrogen density, (f) plasma density, (g) magnetic field strength, (h) B-V angle, (i) B-V plane inclination, (j) He$^+$ density. Panel (k) shows the change due to accounting for the momentum transfer.

## 4. Finding Interstellar Conditions

Finding the interstellar parameters $\pi$ for which the modeled count rates fit the IBEX observations best uses minimization of the following $\chi^2$ expression:

$$\chi^2(\pi) = \sum_{i,j}(c_i - g_i(\pi))(\mathbf{V}^{-1})_{i,j}\left(c_j - g_j(\pi)\right), \qquad (3)$$

where $c_i$ is the observed count rate in data point $i$ and $\mathbf{V}$ is the IBEX data covariance matrix (Section 2). The sum is calculated over all spin axis pointings, ESA steps, and spin angle bins included in the dataset. We minimize the above expression using the series expansion of the modeled count rates in Equation (2). In the minimization, we find the best-fit values of ten interstellar parameters as listed in Table 1, twelve normalization factors $a_{2009}, a_{2010}, \ldots, a_{2020}$, and one parameter $b$ describing the common response function of the instrument, see Equation (1).

The best-fit interstellar parameters are presented in Table 2. The reduced $\chi^2_\nu$ shown in the table is the minimum $\chi^2$ value divided by the number of degrees of freedom. We perform fitting without (rows 1-4) and with (rows 5-8) the momentum transfer effects (the last bracket in Equation (2)). The fits without additional assumptions on the interstellar parameters are listed as unconstrained (rows 2 and 6). Additionally, we consider a case in which only the flow velocity, temperature, and He$^+$ density are sought, while the other interstellar parameters are assumed at their baseline values (selected fixed, rows 1 and 5). Comparison of these two fits shows that the flow velocity and temperature obtained from the fit are moderately, yet significantly, sensitive to the constraints on other interstellar parameters when compared to the uncertainties. The most significant change is in the best-fit He$^+$ density, which changes by a factor of 2. Therefore, this parameter cannot be correctly constrained unless other parameters are also included. Comparison of the unconstrained fits shows that omission of the momentum transfer results in underestimation of the speed by ~0.45 km s$^{-1}$ and overestimation of the temperature by ~660 K. This result confirms the estimation made by Swaczyna et al. (2023b), in which these effects were obtained from comparison of the ISN helium properties far from the Sun with the filtered primary population at the heliopause.

The direct (unconstrained) fit finds the best parameters using only the ISN helium observations. However, the obtained magnetic field strength of 5.4±0.2 µG in the case with momentum transfer is much stronger than 2.93 µG obtained from the analysis of the IBEX ribbon position (Zirnstein et al. 2016). Therefore, we want to check whether we can impose constraints based on other observations that would not significantly change the goodness of fit (see Appendix). Constraining the magnetic field based on the uncertainties reported by Zirnstein et al. (2016), we obtain the best fit magnetic field and plasma density closer to the baseline values (magnetic field constraint, rows 3 and 7 in Table 2). Moreover, a stronger magnetic field and higher plasma density increase the VLISM pressure, resulting in a smaller heliosphere. Therefore, the heliopause distances based on Voyager observations are also a useful constraint (Voyager HP constraint, rows 4 and 8). However, the solar cycle related changes may influence the position of the heliopause (see Appendix for further discussion).

Both constraints result in the best-fit velocities, temperatures, and B-V angles that differ from the unconstrained fit by less than their uncertainties We calculate the mean values from the unconstrained and two constrained fits for these parameters: the flow speed ($v = 26.63±0.17$ km s$^{-1}$), inflow direction in ecliptic coordinates ($\lambda = 255.73°±0.19°$, $\beta = 5.04°±0.15°$), temperature ($T = 7350±110$ K), and B-V plane inclination to the ecliptic plane ($\gamma = 53.7°±0.6°$). These uncertainties account for the standard deviation of the results included in the mean and systematic uncertainty related to the knowledge of instrument boresight relative to the spacecraft coordinate system. The estimated pristine VLISM He$^+$ density is $n_{\text{He}^+} = (9.7±1.2)×10^{-3}$ cm$^{-3}$. The other parameters, such as plasma density, neutral hydrogen density, magnetic field strength, and B-V angle, are strongly affected by the constraints and thus their baseline values should be obtained from other observations, e.g., the position of the IBEX ribbon, the electron density measured by the Voyagers in the outer heliosheath, or PUI observations in the outer heliosphere.

The uncertainties in this table are derived from the curvature of $\chi^2$ around the minimum and are scaled upwards by the square root of the reduced $\chi_\nu^2$. The data collected by IBEX strongly constrain the model parameters, but additional sources of uncertainties from still unaccounted effects are possible. For the number of degrees of freedom in this study, the reduced $\chi_\nu^2$ for a complete model should be between 0.96 and 1.04 at the 99.7% (3σ) confidence level, which indicates that our model ($\chi_\nu^2 = 1.7$–$2.0$) must be incomplete. The high $\chi^2$ value may be because the global heliosphere model that we use is inadequate, there are still unaccounted effects in the filtration of He$^+$ ions, or the ISN helium is not fully equilibrated in the pristine VLISM as suggested by Wood et al. (2019). While the scaling of the uncertainties partially accounts for the related systematic uncertainties, the magnitude of these uncertainties cannot be completely evaluated without fully quantifying these effects. Therefore, the final uncertainties provided here may be underestimated.

The first analysis of IBEX observations showed a clear correlation between the flow speed, longitude, and temperature (Bzowski et al. 2012; Möbius et al. 2012). Figure 3 shows the correlations between the parameters from the covariance matrix analysis obtained from the $\chi^2$ minimization in the unconstrained fit. In addition to the above strongly correlated parameters, the matrix also shows a strong correlation (+0.97) between the magnetic field strength and the plasma density. This correlation means these two parameters cannot be independently constrained using the ISN observations.

Table 2. Best-fit Interstellar Parameters

| | Constraint | Momentum transfer | $\chi^2_\nu$ | $v$ (km s$^{-1}$) | $\lambda$ (°) | $\beta$ (°) | $T$ (K) | $n_{H^0}$ (cm$^{-3}$) | $n_{pl}$ (cm$^{-3}$) | $B$ (µG) | $\alpha$ (°) | $\gamma$ (°) | $n_{He^+}$ (10$^{-3}$ cm$^{-3}$) |
|---|---|---|---|---|---|---|---|---|---|---|---|---|---|
| 1 | Selected fixed | no | 1.992 | 25.65±0.08 | 256.13±0.09 | 5.03±0.02 | 7990±60 | *0.11* | *0.0856* | *2.93* | *39.5* | *52.2* | 5.29±0.13 |
| 2 | Unconstrained | no | 1.737 | 26.26±0.09 | 255.51±0.10 | 5.09±0.02 | 8050±60 | 0.241±0.007 | 0.133±0.009 | 4.50±0.19 | 41.5±1.7 | 54.8±0.5 | 10.4±0.4 |
| 3 | Magnetic field | no | 1.748 | 26.14±0.09 | 255.67±0.10 | 5.11±0.02 | 7980±60 | 0.237±0.005 | 0.084±0.004 | 3.37±0.09 | 38.6±0.7 | 54.0±0.5 | 9.2±0.3 |
| 4 | Voyager HP | no | 1.739 | 26.20±0.09 | 255.57±0.10 | 5.09±0.02 | 8010±60 | 0.245±0.007 | 0.114±0.005 | 4.10±0.14 | 39.0±1.5 | 54.6±0.5 | 10.3±0.4 |
| 5 | Selected fixed | yes | 1.813 | 26.85±0.09 | 255.58±0.09 | 5.02±0.02 | 7560±60 | *0.11* | *0.0856* | *2.93* | *39.5* | *52.2* | 4.74±0.15 |
| 6 | Unconstrained | yes | 1.700 | 26.71±0.09 | 255.64±0.10 | 5.03±0.02 | 7390±60 | 0.179±0.007 | 0.189±0.009 | 5.35±0.18 | 37.0±1.6 | 54.2±0.5 | 10.7±0.4 |
| 7 | Magnetic field | yes | 1.732 | 26.59±0.09 | 255.80±0.10 | 5.07±0.02 | 7330±60 | 0.157±0.005 | 0.122±0.004 | 3.73±0.09 | 36.9±0.7 | 53.0±0.5 | 8.0±0.3 |
| 8 | Voyager HP | yes | 1.706 | 26.60±0.09 | 255.76±0.10 | 5.04±0.02 | 7330±60 | 0.186±0.007 | 0.155±0.006 | 4.62±0.13 | 32.5±1.5 | 53.8±0.5 | 10.4±0.4 |

**Notes.** See the main text for descriptions of the applied constraints. The fixed parameter values for the "selected fixed" constraint are shown in italics. The reported uncertainties include only the uncertainty derived from the $\chi^2$ analysis (see the main text).

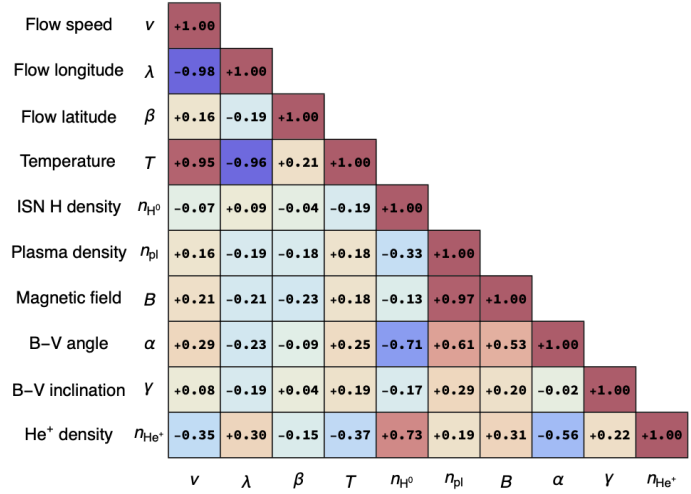

**Figure 3.** Correlation matrix between the fit parameters for the unconstrained case that includes the momentum transfer term. The numbers represent the correlation coefficient between each pair of parameters. Positive (negative) correlations are shown using red (blue) shading.

## 5. Summary

Over the first 12 years of the IBEX mission, the IBEX-Lo instrument detected more than 2.4 million ISN helium atoms allowing for detailed studies of the VLISM near the Sun. Analyses of the IBEX-Lo data allow us to find the VLISM parameters, such as the flow speed, inflow direction, and temperature. In this paper, we account for the charge exchange and elastic collisions in transporting the combined populations of ISN helium from the pristine VLISM to IBEX. We use a global heliosphere model characterized by multiple interstellar parameters to get the flows and temperatures in the outer heliosheath impacting the filtration. We calculate how the integrated ISN helium fluxes depend on these conditions.

We find that the best-fit VLISM flow velocity and temperature change significantly, as estimated by Swaczyna et al. (2023b), if the momentum transfer in elastic and charge exchange collisions is accounted for. The ISN helium observations from IBEX allow for derivation of the flow velocity, temperature, and B-V plane inclination to the ecliptic plane. Our new estimate of the pristine VLISM He$^+$ density $(9.7\pm1.2)\times10^{-3}$ cm$^{-3}$ is consistent with $(8.98\pm0.12)\times10^{-3}$ cm$^{-3}$ derived by Bzowski et al. (2019). However, the ten times larger uncertainty in our analysis points to the importance of a simultaneous analysis of all interstellar parameters. Nevertheless, the reported uncertainties represent only the statistical uncertainty, and unaccounted systematic effects in the modeling or assumptions may exceed these uncertainties.

We showed that our derived best-fit interstellar parameters are susceptible to details of the transport and filtration of ISN helium atoms in the heliospheric boundaries. Moreover, the minimum $\chi^2$ significantly exceeds the expected value, which suggests that the models are still incomplete, and further development of the ISN models is necessary. Future studies may consider ISN helium distribution functions in the pristine VLISM that are not fully thermalized, e.g., kappa distributions or bi-Maxwellian distributions (Swaczyna et al. 2019; Wood et al. 2019) rather than an isotropic Maxwellian distribution used here. This possibility is likely in light of the mixing interstellar cloud medium near the Sun (Swaczyna et al. 2022b). Additionally, the IMAP-Lo instrument on the Interstellar Mapping and Acceleration Probe (IMAP) thanks to new observational capabilities and higher sensitivity (McComas et al. 2018; Sokół et al. 2019; Bzowski et al. 2022, 2023; Schwadron et al. 2022), may allow for more in-depth study of interstellar conditions reflected in the ISN helium fluxes at 1 au.

*Acknowledgments*: This material is based upon work supported by the National Aeronautics and Space Administration (NASA) under grant No. 80NSSC20K0781 issued through the Outer Heliosphere Guest Investigators Program. This work was also partially funded by the IBEX mission as a part of the NASA Explorer Program (80NSSC20K0719) and by IMAP as part of the Solar Terrestrial Probes Program (80GSFC19C0027). M.B. and M.A.K. were supported by the Polish National Science Centre grant 2019/35/B/ST9/01241. The P.S.'s work included in the original submission was carried out at Princeton University, but the revision was prepared at the Space Research Center PAS within a project co-financed by the National Agency for Academic Exchange within Polish Returns Programme (BPN/PPO/2022/1/00017) with a research component funded by the National Science Centre, Poland (2023/02/1/ST9/00004). For the purpose of Open Access, the author has applied a CC-BY public copyright license to any Author Accepted Manuscript (AAM) version arising from this submission.

## Appendix. Constraining the Interstellar Parameters

We use two constraints on sought interstellar parameters from other observations to check the sensitivity of the fitting. The first constraint is defined directly based on the magnetic field vector found in Zirnstein et al. (2016). This study found the best-fit magnetic field using the geometry of the IBEX ribbon. The derived magnetic field strength is 2.93±0.08 μG, while the B-V angle is 39.5°±0.6°. We add the following expression to the minimized $\chi^2$ given in Equation (3):

$$\Delta\chi^2_{\mathrm{mag}}(B, \alpha) = \frac{(B - 2.93)^2}{0.08^2} + \frac{(\alpha - 39.5°)^2}{(0.6°)^2}. \quad (4)$$

This expression adds a penalty term according to the probability distribution of magnetic field parameters obtained from the IBEX ribbon, assuming that this probability is normal.

The second constraint considered in this study is based on the heliopause distances along the trajectories of Voyager 1 and 2 trajectories. We find these positions for the models with the baseline and modified parameter values. After that, the heliopause position is interpolated using the same method as that used to interpolate the modeled count rate shown in Equation (2). We denote these interpolating functions as $r_{\mathrm{HP,V1}}(\boldsymbol{\pi})$ and $r_{\mathrm{HP,V2}}(\boldsymbol{\pi})$ for the heliopause distance along the Voyager 1 and 2 trajectories, respectively. Using these functions, we add the following term to Equation (3):

$$\Delta\chi^2_{\mathrm{Voy}}(\boldsymbol{\pi}) = \frac{\left(r_{\mathrm{HP,V1}}(\boldsymbol{\pi}) - 121.6\right)^2}{10^2} + \frac{\left(r_{\mathrm{HP,V2}}(\boldsymbol{\pi}) - 119.0\right)^2}{10^2}. \quad (5)$$

where 121.6 au and 119 au are the distances at which Voyagers crossed the heliopause (Burlaga & Ness 2014; Burlaga et al. 2019). While the actual crossing positions are known very precisely, we impose relatively large uncertainties of ~10 au on this constraint, which correspond to the approximate time-variation of the heliopause position (Kim et al. 2017; Washimi et al. 2017; Izmodenov & Alexashov 2020). The model used in our study assumes time-independent boundary conditions and thus cannot reproduce this variation.

## References

Burlaga, L. F., & Ness, N. F. 2014, ApJ, 784, 146
Burlaga, L. F., Ness, N. F., Berdichevsky, D. B., et al. 2019, NatAs, 3, 1007
Bzowski, M., Czechowski, A., Frisch, P. C., et al. 2019, ApJ, 882, 60
Bzowski, M., & Heerikhuisen, J. 2020, ApJ, 888, 24
Bzowski, M., & Kubiak, M. A. 2020, ApJ, 901, 12
Bzowski, M., Kubiak, M. A., Czechowski, A., & Grygorczuk, J. 2017, ApJ, 845, 15
Bzowski, M., Kubiak, M. A., Hłond, M., et al. 2014, A&A, 569, A8

Bzowski, M., Kubiak, M. A., Möbius, E., et al. 2012, ApJS, 198, 12
Bzowski, M., Kubiak, M. A., Möbius, E., & Schwadron, N. A. 2022, ApJ, 938, 148
Bzowski, M., Kubiak, M. A., Möbius, E., & Schwadron, N. A. 2023, ApJS, 265, 24
Bzowski, M., Möbius, E., Tarnopolski, S., Izmodenov, V., & Gloeckler, G. 2009, SSRv, 143, 177
Bzowski, M., Sokół, J. M., Tokumaru, M., et al. 2013, in Cross-Calibration of Far UV Spectra of Solar System Objects and the Heliosphere, ed. E. Quémerais, M. Snow, & R.-M. Bonnet (New York: Springer Science + Business Media), 67, http://link.springer.com/chapter/10.1007/978-1-4614-6384-9_3
Bzowski, M., Swaczyna, P., Kubiak, M. A., et al. 2015, ApJS, 220, 28
Elliott, H. A., McComas, D. J., Zirnstein, E. J., et al. 2019, ApJ, 885, 156
Fahr, H. J. 1968, Ap&SS, 2, 474
Fraternale, F., Pogorelov, N. V., & Bera, R. K. 2023, ApJ, 946, 97
Fraternale, F., Pogorelov, N. V., & Heerikhuisen, J. 2021, ApJL, 921, L24
Frisch, P. C., Bzowski, M., Drews, C., et al. 2015, ApJ, 801, 61
Frisch, P. C., Bzowski, M., Livadiotis, G., et al. 2013, Sci, 341, 1080
Fuselier, S. A., Bochsler, P., Chornay, D., et al. 2009, SSRv, 146, 117
Galli, A., Wurz, P., Park, J., et al. 2015, ApJS, 220, 30
Galli, A., Wurz, P., Rahmanifard, F., et al. 2019, ApJ, 871, 52
Galli, A., Wurz, P., Schwadron, N. A., et al. 2017, ApJ, 851, 2
Galli, A., Wurz, P., Schwadron, N. A., et al. 2022, ApJS, 261, 18
Heerikhuisen, J., Zirnstein, E. J., Pogorelov, N. V., Zank, G. P., & Desai, M. 2019, ApJ, 874, 76
Izmodenov, V. V., & Alexashov, D. B. 2015, ApJS, 220, 32
Izmodenov, V. V., & Alexashov, D. B. 2020, A&A, 633, L12
Kim, T. K., Pogorelov, N. V., & Burlaga, L. F. 2017, ApJL, 843, L32
Kubiak, M. A., Bzowski, M., Sokół, J. M., et al. 2014, ApJS, 213, 29
Kubiak, M. A., Swaczyna, P., Bzowski, M., et al. 2016, ApJS, 223, 25
Lallement, R., & Bertaux, J. L. 2014, A&A, 565, A41
Lee, M. A., Kucharek, H., Möbius, E., et al. 2012, ApJS, 198, 10
Lee, M. A., Möbius, E., & Leonard, T. W. 2015, ApJS, 220, 23
Leonard, T. W., Möbius, E., Bzowski, M., et al. 2015, ApJ, 804, 42
McComas, D. J., Allegrini, F., Bochsler, P., et al. 2009, SSRv, 146, 11
McComas, D. J., Angold, N., Elliott, H. A., et al. 2013, ApJ, 779, 2
McComas, D. J., Bzowski, M., Frisch, P., et al. 2015a, ApJ, 801, 28
McComas, D. J., Bzowski, M., Fuselier, S. A., et al. 2015b, ApJS, 220, 22
McComas, D. J., Christian, E. R., Schwadron, N. A., et al. 2018, SSRv, 214, 116
McComas, D. J., Swaczyna, P., Szalay, J. R., et al. 2021, ApJS, 254, 19
Möbius, E., Bochsler, P., Bzowski, M., et al. 2009a, Sci, 326, 969
Möbius, E., Bochsler, P., Bzowski, M., et al. 2012, ApJS, 198, 11
Möbius, E., Bzowski, M., Chalov, S., et al. 2004, A&A, 426, 897
Möbius, E., Bzowski, M., Frisch, P. C., et al. 2015, ApJS, 220, 24
Möbius, E., Kucharek, H., Clark, G., et al. 2009b, SSRv, 146, 149
Patterson, T. N. L., Johnson, F. S., & Hanson, W. B. 1963, P&SS, 11, 767
Pogorelov, N. V., Fichtner, H., Czechowski, A., et al. 2017, SSRv, 212, 193
Rahmanifard, F., Möbius, E., Schwadron, N. A., et al. 2019, ApJ, 887, 217
Ruciński, D., & Bzowski, M. 1995, A&A, 296, 248
Saul, L., Wurz, P., Rodriguez, D., et al. 2012, ApJS, 198, 14
Schwadron, N. A., Möbius, E., Leonard, T., et al. 2015, ApJS, 220, 25
Schwadron, N. A., Möbius, E., McComas, D. J., et al. 2022, ApJS, 258, 7
Sokół, J. M., Kubiak, M. A., Bzowski, M., Möbius, E., & Schwadron, N. A. 2019, ApJS, 245, 28
Sokół, J. M., Kubiak, M. A., Bzowski, M., & Swaczyna, P. 2015, ApJS, 220, 27
Swaczyna, P., Bzowski, M., Fuselier, S. A., et al. 2023a, ApJS, 266, 2
Swaczyna, P., Bzowski, M., Kubiak, M. A., et al. 2015, ApJS, 220, 26

Swaczyna, P., Bzowski, M., Kubiak, M. A., et al. 2018, ApJ, 854, 119
Swaczyna, P., Kubiak, M. A., Bzowski, M., et al. 2022a, ApJS, 259, 42
Swaczyna, P., McComas, D. J., & Schwadron, N. A. 2019, ApJ, 871, 254
Swaczyna, P., McComas, D. J., Zirnstein, E. J., et al. 2020, ApJ, 903, 48
Swaczyna, P., Rahmanifard, F., Zirnstein, E. J., & Heerikhuisen, J. 2023b, ApJ, 943, 74
Swaczyna, P., Rahmanifard, F., Zirnstein, E. J., McComas, D. J., & Heerikhuisen, J. 2021, ApJL, 911, L36
Swaczyna, P., Schwadron, N. A., Möbius, E., et al. 2022b, ApJL, 937, L32
Washimi, H., Tanaka, T., & Zank, G. P. 2017, ApJL, 846, L9
Witte, M., Banaszkiewicz, M., Rosenbauer, H., & McMullin, D. 2004, AdSpR, 34, 61
Wood, B. E., Müller, H.-R., & Möbius, E. 2019, ApJ, 881, 55
Wood, B. E., Müller, H.-R., & Witte, M. 2015, ApJ, 801, 62
Zirnstein, E. J., Heerikhuisen, J., Funsten, H. O., et al. 2016, ApJL, 818, L18